\documentclass[preprint,aps,a4paper,eqsecnum,amsmath,amssymb,nofootinbib]{revtex4}

\usepackage{bbm}

 \newcommand{\klammer}[1]{\left( #1\right)}
 \newcommand{\gklammer}[1]{\left\{ #1\right\}}
 \newcommand{\eckklammer}[1]{\left[ #1\right]}

 \newcommand{\ket}[1]{|#1\rangle}
 \newcommand{\XcalT}[2]{\stackrel{#2}{\mathcal{T}}_{_{\hspace{-.25em}#1}}\hspace{-.15em}}

 \newcommand{\sTP}{\otimes_{\hspace{-.1em}s}}
 \newcommand{\kommut}[2]{\left[ #1,#2\right]}
 \newcommand{\Skommut}[2]{\left[ #1,#2\right]_\pm}
 \newcommand{\eins}{{\mathbbm{1}}}
 \newcommand{\grass}{\boldsymbol{\mathcal{E}}}
 \newcommand{\bi}{\text{i}}
 \newcommand{\sAd}[1]{#1^\sharp}
 \newcommand{\hermit}[1]{#1^\dagger}
 
 \newcommand{\mbs}[1]{#1}

 \newcommand{\spur}[1]{\mathrm{tr}\left\{\, #1 \,\right\}}
 \newcommand{\str}[1]{\mathrm{str}\left\{\, #1 \,\right\}}
 \newcommand{\sTr}[2]{ \mathrm{str}_{_{#1}}\left\{ \, #2\,\right\}}

 \newcommand{\trans}{{^{_T}}}
 \newcommand{\st}{\mathrm{st}}
 \newcommand{\ist}{\mathrm{ist}}
 \newcommand{\sT}[1]{\mathrm{st}_#1}
 \newcommand{\isT}[1]{\mathrm{ist}_#1}

 \newcommand{\MyK}[2]{\stackrel{\scriptscriptstyle{#1}}{K}_{\scriptscriptstyle{#2}}\!}
 
 \newcommand{\MyKTilde}[2]{\stackrel{\scriptscriptstyle{#1}}{\widetilde{K}}_{\scriptscriptstyle{#2}}\!}
 \newcommand{\calA}{\mathcal{A}}
 \newcommand{\calB}{\mathcal{B}}
 \newcommand{\calC}{\mathcal{C}}
 \newcommand{\calD}{\mathcal{D}}
 \newcommand{\calT}{\mathcal{T}}

\begin{document}
\preprint{}
\title{Non-diagonal boundary conditions for $\mathfrak{gl}(1|1)$ super spin
  chains} 

\author{Andr\'e M. Grabinski}

\author{Holger Frahm}

\affiliation{%
Institut f\"ur Theoretische Physik, Leibniz Universit\"at Hannover,
Appelstra\ss{}e 2, 30167 Hannover, Germany}

\date{\today}

\begin{abstract}
  We study a one-dimensional model of free fermions with $\mathfrak{gl}(1|1)$
  supersymmetry and demonstrate how non-diagonal boundary conditions can be
  incorporated into the framework of the graded Quantum Inverse Scattering
  Method (gQISM) by means of \emph{super matrices} with entries from a
  superalgebra.  For super hermitian twists and open boundary conditions
  subject to a certain constraint, we solve the eigenvalue problem for the
  super transfermatrix by means of the graded algebraic Bethe ansatz technique
  (gABA) starting from a fermionic coherent state.  For generic boundary
  conditions the algebraic Bethe ansatz can not be applied.  In this case the
  spectrum of the super transfer matrix is obtained from a functional
  relation.
\end{abstract}

\maketitle

\section{Introduction}
For a long time studies of quantum integrable models in one spatial dimension
have led to important insights into the properties of many body systems and
provided a sound basis for the understanding of the non perturbative phenomena
which arise due to the interplay of interactions and strong quantum
fluctuations in low dimensional systems (see e.g.\ \cite{HUBBARD}).  A special
way to introduce free parameters into these systems is by variation of their
boundary conditions.  Considering all possible classes compatible with the
integrability allows for a complete classification of their low-energy quantum
critical behaviour on one hand but also to study in detail the effect of
embedded impurities and contacts to an environment.  Recently, there has been
increased interest in twisted or non-diagonal boundary conditions which break
certain bulk symmetries of integrable quantum spin chains
\cite{BBOY95,Nepo04,MeRM05,YaNZ06,Galleas08,FrSW08,NiWF09}: although their
hamiltonian is a member of a commuting family of operators the established
algebraic schemes for the computation of the spectrum fail unless additional
constraints to the boundary conditions are in place.  For spin $1/2$ chains
there has been some progress using functional methods, but quite a few open
questions remain. Even less is known for quantum chains with $\mathcal{Z}_2$
grading or higher rank symmetry. Although integrable non-diagonal open
boundary conditions have been constructed
\cite{Zhou96a,Zhou97,GuGR98,Gall07} the solution of the spectral problem is
restricted to diagonal ones so far.

In this paper we study this problem for the simplest possible case of spin
chains with $\mathfrak{gl}(1|1)$ supersymmetry.  Since the corresponding bulk
system describes free spinless fermions on a lattice this should provide a toy
model to investigate in particular the applicability of functional methods to
the solution of the spectral problem.  We begin with a short review of the
graded Quantum Inverse Scattering Method \cite{KuSk82,Kulish85,GoMu98}.  Using
a Grassmann valued super matrix representation of the Yang Baxter algebra,
spin chains subject to twisted periodic boundary conditions can be embedded
into this framework and are solved exactly.  In
Section~\ref{sec:GradedReflectionAlgebra} we construct the
$\mathfrak{gl}(1|1)$ super spin chain with generic open boundary conditions
based on Sklyanin's reflection algebra \cite{Skly88}.  We study the spectrum
of these super spin chains for certain classes of reflection matrices using
the algebraic Bethe ansatz and finally extend this solution to generic
boundaries using functional methods.
\section{Graded Quantum Inverse Scattering Method}
The fundamental objects considered within the framework of the graded Quantum
Inverse Scattering Method (gQISM) are representations $T(v)$ of the
\emph{graded Yang-Baxter algebra} (gYBA) 
\begin{equation}
 R_{12}(u-v)\stackrel{\scriptscriptstyle 1}{T}(u)\stackrel{\scriptscriptstyle 2}{T}(v)=\,
 \stackrel{\scriptscriptstyle 2}{T}(v)\stackrel{\scriptscriptstyle 1}{T}(u)R_{12}(u-v)\, .
 \label{eq:YBA}
\end{equation}
The indices $1$ and $2$ label the linear spaces $V_{1,2}$ into which the
respective operators are embedded by means of the \emph{super tensor product}
$\sTP$, defined through
\begin{equation}
(A \sTP B)(C \sTP D) \equiv (-1)^{p(B)p(C)} AC \sTP BD\, ,
\end{equation}
where $p(X)$ refers to the parity function defined in the appendix. That is,
to be precise 
\begin{equation}
\begin{aligned}
& \stackrel{\scriptscriptstyle 1}{T}(u)\equiv T(u)\sTP\eins\,,\quad \stackrel{\scriptscriptstyle 2}{T}(u)\equiv \eins\sTP T(u)\, ,\\
  R_{12}(u) \equiv R(u)&\sTP\eins\,,\quad 
  R_{23}\equiv\eins\sTP R(u)\quad\text{and}\quad
  R_{13}(u) = P_{23}\,R_{12}(u)\,P_{23}\, .
\end{aligned}
\end{equation}
Here $P_{ij}$ is the \emph{graded permutation operator} that interchanges two
spaces $V_i$ and $V_j$ according to $P(x\sTP y) \equiv (-1)^{p(x)p(y)}(y\sTP
x)$. The $R$-matrix is subject to the consistency condition
\begin{equation}
 R_{12}(u-v) R_{13}(u) R_{23}(v)=\,
 R_{23}(v) R_{13}(u) R_{12}(u-v)\, ,
 \label{eq:YBE}
\end{equation}
known as \emph{Yang-Baxter equation} (YBE). As a consequence one obtains local
representations $L_{0j}(u)\equiv R_{0j}(u)$ of the gYBA by a graded embedding
of the $R$-matrix.  These \emph{Lax-operators} $L_{0j}(u)$ act on an auxiliary
space $V_0$, whereas their entries act on the $j$-th quantum space $V_j$. Due
to its \emph{comultiplication} property, the gYBA allows for the construction
of global representations as products of Lax-operators.  This results in a
particular representation on the auxiliary space and the tensor product of the
quantum spaces $V_q = V_1\sTP V_2\sTP\cdots\sTP V_N$, the
\emph{monodromy matrix}
\begin{equation}
 T(u)\equiv L_{0N}(u) L_{0,N-1}(u)\dots L_{01}(u)\, .
 \label{eq:MonodromieMatrix}
\end{equation}
Taking the supertrace (\ref{eq:superTrace}) of this monodromy matrix, we
obtain the \emph{super transfermatrix} $\tau(u) = \str{T(u)}$ which generates
a set of commuting operators on $V_q$. In particular, it is related to an
integrable hamiltonian with periodic boundary conditions defined by
$H=\partial_u\ln\tau(u)|_{u=0}$.

For the $\mathfrak{gl}(1|1)$ supersymmetric representations of the gYBA
considered here, this construction leads to a model of free spinless fermions
on a one-dimensional lattice with $N$ sites. In the case of periodic boundary
conditions the hamiltonian reads
\begin{equation}
 H=\sum_{j=1}^N H_{j,j+1}\quad ,\quad H_{j,j+1} \equiv \klammer{\hermit{c}_j c_{j+1}^{\phantom{\dagger}} + \hermit{c}_{j+1} c_j^{\phantom{\dagger}}}-n_j^{\phantom{\dagger}}-n_{j+1}^{\phantom{\dagger}}+1\, .
 \label{eq:PBCHamiltonian}
\end{equation}
The corresponding $R$-matrix (cf. \cite{GoMu98}) is 
\begin{equation}
 R(u) = 
 \begin{pmatrix}
  u+1 & ~ & ~ &  ~  \\
   ~  & u & 1 &  ~  \\
   ~  & 1 & u &  ~  \\
   ~  & ~ & ~ & u-1
 \end{pmatrix}\quad\curvearrowright\quad
 \check{R}(u) \equiv P\,R(u) =
 \begin{pmatrix}
  1+u & ~ & ~ &  ~  \\
   ~  & 1 & u &  ~  \\
   ~  & u & 1 &  ~  \\
   ~  & ~ & ~ & 1-u
 \end{pmatrix}
 \label{eq:gl11RMatrix}
\end{equation}
and a graded embedding yields
\begin{equation}
 L_{0j}(u)\equiv R_{0j}(u)=\begin{pmatrix}
                            u+e_j,_1^{~1} & e_j,_2^{~1} \\[.3em]
                            e_j,_1^{~2}   & u-e_j,_2^{~2}
                           \end{pmatrix}
                          =\begin{pmatrix}
                            u+\bar{n}_j & \hermit{c}_j \\
                            c_j         & u-n_j
                           \end{pmatrix}\, .
 \label{eq:SuperLaxe}
\end{equation}
Generally we will define the hamiltonian density in terms of the checked $R$-matrix via $H_{ij} \equiv \partial_u\check{R}_{ij}(u)|_{u=0}$.
\subsection{Super hermitian twists}
\label{sec:supertwist}
The simplest generalization of periodic boundary conditions are twists. They
can easily be incorporated into the above scheme by making use of the
comultiplication property again. Let the \emph{twist matrix} $K$ be a
representation of the gYBA on the auxiliary space.  Then $K \cdot T(u)$ is
another global representation producing the super transfermatrix
\begin{equation}
 \tau(u) = \str{K\cdot T(u)} =\str{K\, L_{0N}(u) L_{0,N-1}(u) \dots L_{01}(u)
 } \, , 
 \label{eq:TwistTransfer}
\end{equation}
which results in a modified hamiltonian on $V_q$ which contains a boundary
term
\begin{equation}
 H_{\text{twist}} = \sum_{j=1}^{N-1} H_{j,j+1} + K^{-1}_N H_{N1}^{\phantom{1}} K^{\phantom{1}}_N\, .
\end{equation}
As a specific twist matrix we choose
\begin{equation}
 \mbs{K} = \begin{pmatrix}
       a               & d\grass \\
       \sAd{(d\grass)} & b       \\
     \end{pmatrix}\quad a,b\in\mathbb{R}\; ,\; d\in\mathbb{C}\, ,
 \label{eq:sh_K}
\end{equation}
where $\grass$ is the sole generator of $\mathbb{C}\boldsymbol{G}_1$ (see
Appendix \ref{sec:grassm}). Notice that this is the most general
$\mathbb{C}\boldsymbol{G}_1$ super matrix being hermitian with respect to the
operation (\ref{eq:superAdjoint}). Taking into account the properties of
Grassmann numbers, $K$ can be diagonalized by a \emph{super unitary}
transformation
\begin{equation}
 \mbs{U} = \frac{1}{a-b}
            \begin{pmatrix}
             \bi(a-b) & d\sAd\grass \\
             d^*\grass   & \bi(a-b)\\
            \end{pmatrix}\quad\curvearrowright\quad
 \hermit{\mbs{U}}\mbs{U}=\mbs{U}\hermit{\mbs{U}}=\eins\, ,
  \label{eq:TransU_K}
\end{equation}
such that
\begin{equation}
 \widetilde{K} \equiv \hermit{\mbs{U}}\mbs{K}\mbs{U}= \begin{pmatrix}
                               a & \quad \\
                               \quad & b
                             \end{pmatrix}\, .
\end{equation}
The Lax-operators (\ref{eq:SuperLaxe}) are super matrices over the algebra
described in Appendix \ref{sec:gl11}, hence the comultiplication
(\ref{eq:TwistTransfer}) will lead to products between fermionic operators
(\ref{eq:gl11fermi}) and Grassmann numbers. For homogeneous elements
$C\in\mathcal{F}$ and $G\in\mathbb{C}\boldsymbol{G}_{\mathcal{N}}$ we define
\begin{equation}
 \Skommut{C}{G}	 = 0 \quad\text{and}\quad p(CG)=p(GC)\equiv p(G)+p(C) \mod 2\, .
\end{equation}
In the periodic case (\ref{eq:PBCHamiltonian}), the spectrum can be obtained
by means of the graded algebraic Bethe ansatz (gABA) with the
Fock-vacuum as a reference state.  For diagonal (or upper triangular) twist
matrix $K$ the Fock-vacuum would still provide a suitable reference state for
the gABA.  For more general twists a different pseudo vacuum has to be used.

Using the cyclicity of the supertrace we rewrite the super transfermatrix
(\ref{eq:TwistTransfer}) as 
\begin{equation}
 \tau(u) = \str{\widetilde{K}\, \widetilde{L}_{0N}(u) \widetilde{L}_{0,N-1}(u) \dots \widetilde{L}_{01}(u) }\, ,
 \label{eq:TwistTransferTilde}
\end{equation}
with \emph{transformed} Lax-operators
\begin{equation}
 \widetilde{L}_{0j}(u) \equiv \hermit{U}\, L_{0j}(u)\,U =
 \begin{pmatrix}
  u+\bar{n}_j-\frac{d^*}{a-b}\sAd{\grass}\hermit{c}_j+\frac{d}{a-b}\grass c_j & \hermit{c}_j - \frac{d}{a-b}\grass \\[.7em]
  c_j - \frac{d^*}{a-b}\sAd{\grass} & u - n_j -\frac{d^*}{a-b}\sAd{\grass}\hermit{c}_j+\frac{d}{a-b}\grass c_j
 \end{pmatrix}\, .
\end{equation}
By means of a super unitary transformation on the quantum space $V_j$ the
Lax-operator $\widetilde{L}_{0j}(u)$ can be written in the form
(\ref{eq:SuperLaxe}): setting 
\begin{equation}
 \rho\equiv\frac{d^*}{a-b}\sAd{\grass}\quad\curvearrowright\quad\sAd{\rho}=\frac{d}{a-b}\grass
\end{equation}
we define unitary operators
\begin{equation}
\label{eq:QuantenTrafo}
\begin{aligned}
 Q_j &\equiv \eins + \sAd{\rho}c_j + \rho\hermit{c}_j
      = e^{\rho \hermit{c}_j + \sAd{\rho}c_j}
 \quad\curvearrowright\quad 
 \hermit{Q}_j = \eins - \rho\hermit{c}_j - \sAd{\rho}c_j
      = e^{-(\rho \hermit{c}_j + \sAd{\rho}c_j)}
\end{aligned}
\end{equation}
that map the fermionic creation and annihilation operators according to
\begin{equation}
\begin{aligned}
 \tilde{c}_j &= \hermit{Q}_j c_j Q_j= c_j - \rho
 \quad \text{and}\quad
 \hermit{\tilde{c}}_j = \hermit{Q}_j \hermit{c}_j Q_j=\hermit{c}_j-\sAd{\rho}
\\
 \curvearrowright\quad \tilde{n}_j \equiv \hermit{\tilde{c}}_j\tilde{c}_j &= n_j+\rho \hermit{c}_j -\sAd{\rho}c_j\,,\quad
 \tilde{\bar{n}}_j \equiv \tilde{c}_j\hermit{\tilde{c}}_j = 1-\tilde{n}_j =
 1-n_j-\rho\hermit{c}_j +\sAd{\rho}c_j\, .
\end{aligned}
\end{equation}
In terms of these new fermionic creation and annihilation operators we obtain
\begin{equation}
 \widetilde{L}_{0j}(u) =\begin{pmatrix}
                          u+\tilde{\bar{n}}_j & \hermit{\tilde{c}}_j \\
                            \tilde{c}_j       & u-\tilde{n}_j
                        \end{pmatrix}\, .
 \label{eq:TrafoLaxe}
\end{equation}
After this transformation the gABA can be applied with the new Fock vacuum
\begin{equation}
 \ket{\tilde{0}} = e^{-\rho\sum_{j=1}^{N}\hermit{c}_j}\ket{0}\,
 \label{eq:PseudoVakuum}
\end{equation}
as the reference state.  Note that the local Fock vacua
\begin{equation}
\ket{\tilde{0}_j} = \hermit{Q}_j \ket{0_j} = \ket{0_j} - \rho \ket{1_j}\,
\end{equation}
are \emph{fermionic coherent states}, i.e.\ eigenstates of the annihilation
operator $c_j\ket{\tilde{0}_j} = \rho \ket{\tilde{0}_j}$.
\section{Graded reflection algebra}
\label{sec:GradedReflectionAlgebra}
We will now extend Sklyanin's formalism for the treatment of integrable
systems with open boundary conditions \cite{Skly88} in a way that makes it
applicable to supersymmetric models. Following \cite{Gonz94,BGZZ98}, for a
given $R$-matrix we introduce two associative superalgebras $\XcalT{-}{}$ and
$\XcalT{+}{}$, subject to the \emph{graded reflection equation}
\begin{eqnarray}
  &&R_{12}(u-v)\XcalT{-}{1}(u)R_{21}(u+v)\XcalT{-}{2}(v)\notag\\
  &&\phantom{R_{12}(u)}=\,\XcalT{-}{2}(v)R_{12}(u+v)\XcalT{-}{1}(u)R_{21}(u-v)
  \label{eq:RA1}
\end{eqnarray}
and to the \emph{dual graded reflection equation}
\begin{eqnarray}
  &&R_{21}^{\sT{1}\isT{2}}(v-u)\XcalT{+}{1}^{\hspace{-0.2em}\sT{1}}(u)\widetilde{R}_{12}(-u-v)\XcalT{+}{2}^{\hspace{-0.2em}\isT{2}}(v)\notag\\
  &&\phantom{R_{21}^{\sT{1}\isT{2}}(v)}=\,\XcalT{+}{2}^{\hspace{-0.2em}\isT{2}}(v)\bar{R}_{21}(-u-v)\XcalT{+}{1}^{\hspace{-0.2em}\sT{1}}(u)R_{12}^{\sT{1}\isT{2}}(v-u)
  \label{eq:RA2}
\end{eqnarray}
respectively, whereas the new matrices $\widetilde{R}$ and $\bar{R}$ are related to the $R$-matrix via
\begin{eqnarray}
 \widetilde{R}_{12}^{\sT{2}}(-u-v)R_{21}^{\sT{1}}(u+v)&=&\eins\label{eq:RA_def1}\quad \text{and}\\
 \bar{R}_{21}^{\isT{1}}(-u-v)R_{12}^{\isT{2}}(u+v)&=&\eins\label{eq:RA_def2}\, .
\end{eqnarray}
Moreover the $R$-matrix (\ref{eq:gl11RMatrix}) satisfies the unitarity
condition $R_{12}(u-v)R_{21}(v-u)\sim\eins$.
Under these conditions it is possible to show that the super transfermatrices
\begin{equation}
 \tau(u)\equiv\str{\XcalT{+}{}(u)\XcalT{-}{}(u)}	
 \label{eq:RA_supertrans}
\end{equation}
provide a family of commuting operators, i.e. $\kommut{\tau(u)}{\tau(v)}=0$,
$\forall\, u,v\in\mathbb{C}$.

Now open boundary conditions can be described by two auxiliary space matrices
$K_-(u)$ and $K_+(u)$ satisfying the reflection equations
(\ref{eq:RA1}) and (\ref{eq:RA2}). Up to normalization, the restriction
to $\mathbb{C}\boldsymbol{G}_1$ essentially\footnote{Constant matrices of the form $K_{\pm}=(K_{\pm}(u)-\eins)/u$ can be employed as well.} yields solutions
\begin{equation}
 K_{\pm}(u) = \eins+u\,\begin{pmatrix}
                   a_{\pm}       & b_{\pm}\,\grass \\
                   f_{\pm}\,\sAd{\grass} & -a_{\pm}
                 \end{pmatrix}\hspace{5em}\phantom{.}\label{eq:TypeI}
\end{equation}
with complex coefficients $a_{\pm},b_{\pm}$ and $f_{\pm}$.

Let $T(u)$ be a representation of the gYBA (\ref{eq:YBA}). Then $T(u)K_-(u)T^{-1}(-u)$ is a further representation of the graded reflection algebra $\XcalT{-}{}$ and we have
\begin{equation}
 \tau(u)=\str{K_+(u)T(u)K_-(u)T^{-1}(-u)}\, .
 \label{eq:Supertrans}
\end{equation}
The $R$-matrix is regular, that is $R(0) = P$, and for convenience let us
choose the normalization such that $K_-(0)=\eins$. Since $K_+(0)$ has a
vanishing supertrace we compute the second derivative of the super
transfermatix (\ref{eq:Supertrans}) and -- bearing in mind that the $R$-matrix
(\ref{eq:gl11RMatrix}) complies with the unitarity condition only up to
normalization -- find
\begin{equation}
 \left.\frac{d^2}{du^2}\tau(u)\right|_{u=0}=8\left[1+a_{+}\right]H
 \label{eq:noSTRsecondDeriv}
\end{equation}
with the open chain hamiltionian
\begin{equation}
 H = \sum_{j=1}^{N-1}H_{j,j+1}+\frac{1}{2}\left.\frac{d}{du}\stackrel{\scriptscriptstyle 1}{K}_-\!(u)\right|_{u=0}
       +\frac{1}{2(1+a_{+})}\left.\frac{d}{du}\stackrel{\scriptscriptstyle N}{K}_+\!(u)\right|_{u=0}\, .
\label{eq:OBCHamiltonian2}
\end{equation}
Now we may address the question of what type of boundary terms the matrices
$K_-$ and $K_+$ do generate, i.e. in what way such boundary
conditions affect the hamiltonian of the given model. Using the expressions
(\ref{eq:TypeI}) explicitly, the hamiltonian (\ref{eq:OBCHamiltonian2}) can be
written as
\begin{equation}
 H = \sum_{j=1}^{N-1} H_{j,j+1} + \frac{1}{2}\begin{pmatrix}
                                          a_- & d_-\grass \\
                                          f_-\sAd{\grass} & -a_-
                                         \end{pmatrix}_{\!\!1}
    +\frac{1}{2(1+a_+)}\begin{pmatrix}
                             a_+ & d_+\grass \\
                             f_+\sAd{\grass} & -a_+
                            \end{pmatrix}_{\!N}\ .
\end{equation}
In using standard representations of (\ref{eq:gl11fermi}) and by exploiting (\ref{eq:MultiplGrass}) we can express the two matrices from the latter equation by elements of the combined superalgebra. The first matrix yields
\begin{eqnarray}
  \begin{pmatrix}
   a_- & d_-\grass \\
   f_-\sAd{\grass} & -a_-
  \end{pmatrix}
 &=&\begin{pmatrix}
   a_- &          \\
            & a_-
  \end{pmatrix}
 +d_- \begin{pmatrix}
            0 & \grass \\
            0 & 0
           \end{pmatrix}
 +f_- \begin{pmatrix}
            0            & 0 \\
            \sAd{\grass} & 0
           \end{pmatrix}\\[1em]
 &=&a_- \begin{pmatrix}
            1 & 0 \\
            0 & -1
           \end{pmatrix}
 +d_- \begin{pmatrix}
            \grass &        \\
                   & -\grass
           \end{pmatrix}
           \begin{pmatrix}
            0 & 1 \\
            0 & 0
           \end{pmatrix}
 +f_- \begin{pmatrix}
            \sAd{\grass} &              \\
                         & -\sAd{\grass}
           \end{pmatrix}
           \begin{pmatrix}
            0  & 0 \\
            -1 & 0
           \end{pmatrix}\qquad\quad\\[1em]
 &=&a_- (\eins-2n) + d_-\grass\, c-f_-\sAd{\grass}\hermit{c}
\end{eqnarray}
and after repeating this procedure for the second matrix, the entire hamiltonian reads
\begin{eqnarray}
 H = \sum_{j=1}^{N-1} H_{j,j+1}
     &+&\frac{1}{2}\eckklammer{a_- -2a_- n_1+d_-\grass c_1 - f_- \sAd{\grass}\hermit{c}_1}\notag \\
     &+&\frac{1}{2(1+a_+)}\eckklammer{a_+ -2a_+ n_N+ d_+\grass c_N - f_+ \sAd{\grass}\hermit{c}_N}\, .
 \label{eq:gl11GrassmannHamiltonian}
\end{eqnarray}
We point out that the non-diagonal boundary terms, which do not preserve the
particle number, are Grassmann valued (i.e. $\sim \grass$).  Such terms may
arise, e.g., in the description of the system coupled to a fermionic
environment after integrating out the bath degrees of freedom.
\section{Graded algebraic Bethe ansatz}
In this section we show how the spectral problem for the hamiltonian
(\ref{eq:gl11GrassmannHamiltonian}) can be solved by means of a graded
algebraic Bethe ansatz. For notational convenience we set $\calT(u) \equiv
T(u) K_{-}(u) T^{-1}(-u)$ and consider $\calT(u)$ as a $2\times 2$-matrix
\begin{equation}
 \calT(u)\equiv
       \begin{pmatrix}
        \calA(u) & \calB(u) \\
        \calC(u) & \calD(u)
       \end{pmatrix}
\end{equation}
on the auxiliary space. The reflection equation (\ref{eq:RA1}) gives commutation relations between the quantum space operators $\calA(u), \calB(u), \calC(u)$ and $\calD(u)$ of which the following three are of particular interest
\begin{subequations}
\begin{eqnarray}
 \calB(u)\calB(v)&=&\frac{1-u+v}{1+u-v}\,\calB(v)\calB(u)\, ,\label{eq:VertRelaA}\\[1em]
 \calA(u)\calB(v)&=&\frac{(1-u+v)(v+u)}{(1+u+v)(v-u)}\,\calB(v)\calA(u)+\frac{1}{1+u+v}\,\calB(u)\gklammer{\frac{u+v}{u-v}\,\calA(v)-\calD(v)}\, ,\qquad\qquad\label{eq:VertRelaB}\\[1em]
 \calD(u)\calB(v)&=&\frac{(1-u+v)(v+u)}{(1+u+v)(v-u)}\,\calB(v)\calD(u)+\frac{1}{1+u+v}\,\calB(u)\gklammer{\frac{u+v}{u-v}\,\calD(v)-\calA(v)}\, .\qquad\qquad\label{eq:VertRelaC}
\end{eqnarray}
\end{subequations}
Let $\ket{0}$ be a pseudo-vacuum upon which $\calT(u)$ acts as an upper triangular matrix, i.e.
\begin{equation}
 \calT(u)\ket{0} = \begin{pmatrix}
                     \calA(u)\ket{0} & \calB(u)\ket{0} \\
                     \calC(u)\ket{0} & \calD(u)\ket{0}
                   \end{pmatrix}
                 = \begin{pmatrix}
                     \alpha(u)\ket{0} & *\neq 0         \\
                             0        & \delta(u)\ket{0}
                   \end{pmatrix}\, .
\end{equation}
Here $\alpha(u)$ and $\delta(u)$ are scalar functions, called
\emph{parameters}, that are to be determined later on. They are eigenvalues to
$\calA(u)$ and $\calD(u)$ for the eigenstate $\ket{0}$.
\subsection{Diagonal boundary conditions}
\label{sec:DiagonalABA}
We begin by considering \emph{diagonal} boundary matrices $K_-$ and
$K_+$, i.e.\ 
\begin{equation}
 K_-(u) = \begin{pmatrix}
                1+u a_- & \\
                             & 1-u a_-
               \end{pmatrix}
 \quad\text{and}\quad
 K_+(u) = \begin{pmatrix}
                1+u a_+ & \\
                             & 1-u a_+
               \end{pmatrix}\, .
\end{equation}
This yields the super transfermatrix
\begin{equation}
 \tau(u) = \str{K_+(u)\calT(u)} = (1+u a_+)\calA(u)-(1-u a_+)\calD(u)\, . 
\end{equation}
Using the commutation relations (\ref{eq:VertRelaA}) to
(\ref{eq:VertRelaC}) we find $\calB(v_1)\dots\calB(v_M)\ket{0}$ to be an
eigenstate of $\tau(u)$ with eigenvalue
\begin{equation}
 \Lambda(u) = \eckklammer{\prod_{\ell = 1}^M \frac{(1-u+v_\ell)(v_\ell+u)}{(1+u+v_\ell)(v_\ell-u)}}
              \klammer{(1+u a_+)\alpha(u)\stackrel{\phantom{\prod}}{-}(1-u a_+)\delta(u)}\, ,
 \label{eq:EigenwertSupertransfermatrix}
\end{equation}
provided that the \emph{Bethe ansatz equations}
\begin{equation}
 \frac{\alpha(v_j)}{\delta(v_j)}=\frac{1-a_{+}\,v_j}{1+a_{+}\,v_j}
\end{equation}
are satisfied.  Here the functions $\alpha(u)$ and $\delta(u)$ are
obtained from the action of $\mathcal{T}(u)$ on the Fock vacuum $\ket{0}$ 
\begin{equation}
\label{eq:LaxIteration}
\begin{aligned}
 \calT(u)\ket{0} &= T(u)\MyK{0}{-}(u)T^{-1}(-u)\ket{0}
%\\
%                  &= L_{0N}(u)\dots L_{01}\MyK{0}{-}(u)L^{-1}_{01}(-u)\dots
%                  L^{-1}_{0N}(-u)\ket{0}\\ 
%                  &= \klammer{\frac{1}{1-u^2}}^N L_{0N}(u)\dots
%                  L_{01}\MyK{0}{-}(u) L_{01}(u)\dots
%                  L_{0N}(u)\ket{0}\\ 
                 &= \begin{pmatrix}\alpha(u) & \calB(u)\\ 0 &
                   \delta(u)\end{pmatrix}\ket{0}\, .
\end{aligned}
\end{equation}
Using (\ref{eq:MonodromieMatrix}) and (\ref{eq:SuperLaxe}) we find
\begin{equation}
\begin{aligned}
 \alpha(u) &= \klammer{\frac{1}{1-u^2}}^N (1+u a_-)[u+1]^{2N}\\
 \delta(u) &= \klammer{\frac{1}{1-u^2}}^N \gklammer{(1-u a_-)u^{2N}+(1+u a_-)\frac{u^{2N}}{1+2u}\klammer{\eckklammer{\frac{u+1}{u}}^{2N}-1}}\, .
\end{aligned}
\end{equation}
Therefore the Bethe ansatz equations
\begin{equation}
\label{eq:BAE}
  \left(\frac{v_j+1}{v_j}\right)^N =
   \frac{1-a_+ v_j}{1+a_+ (v_j+1)} \, \frac{1-a_- (v_j+1)}{1+a_- v_j}
\end{equation}
determine the quantization of single particle momenta of the free fermions due
to the boundary conditions.

Finally, we find an explicit expression for the operators $\calB(u)$, that
generate eigenstates of the super transfermatrix:
\begin{eqnarray}
\begin{aligned}
 \calB(u) = \klammer{\frac{u}{1-u}}^{N}\frac{2}{2u+1}\sum_{\ell=1}^{N}&\left\{
 \eckklammer{1+ua_{-}}\klammer{\frac{u+1}{u}}^{j-1}\right.\\
 &\hspace{2em}\left.+\eckklammer{\frac{u}{u+1}-a_{-}}\klammer{\frac{u}{u+1}}^{j-1}\right\}
 \hermit{c}_{\ell}\, .
\end{aligned}
 \label{eq:BOperatorExplizit}
\end{eqnarray}
%
%%%%%%%%%%%%%%%%%%%%%%%%%%%%%%%%%%%%%%%%%%%%%%%%%%%%%%%%%%%%%%%%%%%%%%
\subsection{Quasi-diagonal boundary conditions}
\label{sec:NonDiagonalBethe}
Application of the graded Bethe ansatz for \emph{non}-diagonal boundary
matrices is only possible when a suitable reference state can be found.  Here
we consider a super hermitian left boundary matrix $K_+$ 
\begin{equation}
 K_{+}(u) = \eins+u\,\begin{pmatrix}
                          a_+               & d_+\grass \\
                          d^*_+\sAd{\grass} & -a_+
                         \end{pmatrix}
                         \quad\text{with}\quad a_+\in\mathbb{R}\quad
                         \mathrm{und}\quad d_+\in\mathbb{C}\, ,  
 \label{eq:KPlusABA}
\end{equation}
which is diagonalized by the super unitary transformation
\begin{equation}
 U = \frac{1}{2a_+}\begin{pmatrix}
                         2\bi a_+    & d_+ \sAd{\grass} \\
                         d^*_+\grass & 2\bi a_+
                        \end{pmatrix}
\,,\qquad
 \widetilde{K}_+(u) = \hermit{U}K_+(u) U
                         = \begin{pmatrix}
                            1+u a_+ &             \\
                                         & 1-u a_+
                           \end{pmatrix}\, .
 \label{eq:SuperUniTrafo}
\end{equation}
Now we proceed as in Section \ref{sec:supertwist}: the transformation $U$
leaves the Lax-operators shape-invariant, and we find
\begin{equation}
 \widetilde{L}_{0j}(u) = \hermit{U}\, L_{0j}(u)\,U
                       = \begin{pmatrix}
                          u+\tilde{\bar{n}}_j & \hermit{\tilde{c}}_j \\
                            \tilde{c}_j       & u-\tilde{n}_j
                         \end{pmatrix}\, ,
\end{equation}
where $\tilde{c}_j = c_j-\rho$ and $\hermit{\tilde{c}}_j = \hermit{c}_j -
\sAd{\rho}$; but now we have
\begin{equation}
 \rho\equiv\frac{d^*_+}{2 a_+}\sAd{\grass}\quad\curvearrowright\quad\sAd{\rho}=\frac{d_+}{2 a_+}\grass\, .
 \label{eq:NewRho}
\end{equation}
Due to the cyclicity of the supertrace the super transfermatrix can be written
as
\begin{equation}
 \tau(u) = \sTr{0}{\MyK{0}{+}(u)\calT(u)} = \sTr{0}{\MyKTilde{0}{+}(u)\widetilde{\calT}(u)}\, ,
\end{equation}
where
\begin{equation}
 \widetilde{\calT}(u) =
 \widetilde{T}(u)\MyKTilde{0}{-}(u)\widetilde{T}^{-1}(-u)
 \equiv \begin{pmatrix} \widetilde{\calA}(u) &
                        \widetilde{\calB}(u) \\ \widetilde{\calC}(u) &
                        \widetilde{\calD}(u) \end{pmatrix}\, .
\end{equation}
Here we have introduced $\widetilde{T}(u) = \widetilde{L}_{0N}(u)\dots
\widetilde{L}_{01}(u)$ and $\widetilde{K}_-$ is the transformed right boundary
matrix (\ref{eq:TypeI})
\begin{equation}
\label{eq:Ktransf}
 \widetilde{K}_-(u) = \hermit{U}K_-(u) U
                         = \frac{1}{a_+}
                           \begin{pmatrix}
                            a_{+}(1+u a_-)                                & (a_{+}d_{-}-d_{+}a_{-})u\grass \\
                            (a_{+}f_{-}-d^*_{+}a_{-})u\sAd{\grass} & a_{+}(1-u a_-)
                           \end{pmatrix}\,.
\end{equation}
Now, choosing the parameters in (\ref{eq:Ktransf}) to satisfy the constraint
\begin{equation}
\label{eq:LRconstraint}
   a_{+}f_{-}=d^*_{+}a_{-}\,,
\end{equation}
the transformed boundary matrix $\widetilde{K}_-$ is upper triangular and the
graded algebraic Bethe ansatz can be performed again with a pseudo vacuum
constructed from the fermionic coherent state (\ref{eq:PseudoVakuum}) by using
the definition (\ref{eq:NewRho}) for $\rho$ (see Ref.~\onlinecite{MeRM05} for
a similar approach in the ungraded case).  Furthermore, since the transformed
quantum space operators $\widetilde\calA(u), \widetilde\calB(u),
\widetilde\calC(u)$ and $\widetilde\calD(u)$ obey the same fundamental
commutation relations (\ref{eq:VertRelaA}) to (\ref{eq:VertRelaC}) as their
original counterparts, the Bethe ansatz equations (\ref{eq:BAE}) remain
unchanged.

Compared to the diagonal case we find that the addition of non-diagonal
boundary parameters subject to the constraint (\ref{eq:LRconstraint}) does not
affect the eigenvalues of the super transfermatrix: the energy spectrum of the
chain is determined by the diagonal parameters $a_\pm$ of the boundary
matrices alone.
The Bethe states are generated by the action of the operator $\widetilde\calB$
on the new pseudo vacuum.  Due to the unitary transformation it contains a
Grassmann valued shift
\begin{equation}
\begin{aligned}
 \widetilde\calB(u) =
 &\klammer{\frac{u}{1-u}}^{N}\frac{2}{2u+1}\sum_{\ell=1}^{N}\left\{ 
 \eckklammer{1+ua_{-}}\klammer{\frac{u+1}{u}}^{j-1}\right.\\
 &\hspace{3em}\left.+\eckklammer{\frac{u}{u+1}-a_{-}}
 \klammer{\frac{u}{u+1}}^{j-1}\right\} 
 \hermit{\tilde{c}}_{\ell} + \klammer{\frac{u}{1-u}}^N
 \klammer{d_{-}-\frac{d_{+}}{a_{+}}a_{-}} u \grass\, . 
\end{aligned} 
 \label{eq:BSchlangeExplizit}
\end{equation}
Therefore, the Bethe states $\widetilde\calB(v_1)\dots
\widetilde\calB(v_M)\ket{\widetilde0}$ are linear combinations of states with
up to $M$ particles added to the coherent state Fock vacuum
(\ref{eq:PseudoVakuum}).

%%%%%%%%%%%%%%%%%%%%%%%%%%%%%%%%%%%%%%%%%%%%%%%%%%%%%%%%%%%%%%%%%%%%%%
\subsection{Generic boundary conditions: functional relations}
Finally, we want to address the question to what extent the spectral problem
of the $\mathfrak{gl}(1|1)$-model can be solved if we choose more general
boundary matrices than those allowed by the constraint
(\ref{eq:LRconstraint}).  In this case a reference state suitable for the
application of the gABA is not available.

In the case of spin $1/2$ chains without grading this question has been
addressed by exploiting certain functional relations obeyed by the eigenvalues
of the transfer matrix as a consequence of integrability of the model (see
e.g. \cite{BBOY95,Nepo04,YaNZ06,FrSW08}).  To obtain such a functional
relation for the model considered here we begin with the representation
(\ref{eq:EigenwertSupertransfermatrix}) of the eigenvalues in terms of roots
of the Bethe equations.  Note that only the eigenvalues of the boundary
matrices enter this expression in the cases studied above.

Let $k_\pm^{1,2}$ be the eigenvalues of the boundary matrices $K_\pm(u)$, then
(\ref{eq:EigenwertSupertransfermatrix}) can be rewritten as a functional
relation for an unknown function $q(u)$
\begin{equation}
 \Lambda(u) = \frac{q(u-1)}{q(u)}\,f(u)
 \label{eq:GrTQGleichung}
\end{equation}
where $f(u)$ is a known function:
\begin{equation}
\begin{aligned}
  f(u) &\equiv k_{+}^{1} \alpha(u) - k_{+}^{2} \delta(u)\\
  &= k_{+}^{1} \klammer{\frac{1}{1-u^2}}^N k_{-}^{1}
  [u+1]^{2N} \\
  &\phantom{=~} - k_{+}^{2} \klammer{\frac{u^2}{1-u^2}}^N
  \gklammer{k_{-}^{2} + \frac{k_{-}^{1}}{1+2u}
    \klammer{\eckklammer{\frac{u+1}{u}}^{2N}-1}}\,.
    \end{aligned}
\label{eq:funcF}
\end{equation}
By construction $\Lambda(u)$ is a polynomial in $u$.  Therefore Eq.\
(\ref{eq:GrTQGleichung}) has to be complemented with the condition that its
RHS is analytic.  In particular the residues at the zeroes of the unknown
function $q(u)$ have to vanish.  With a polynomial ansatz
\begin{equation}
 q(u) \equiv \prod_{\ell = 1}^{M}{(-u-1-v_{\ell})(u-v_{\ell})}\, ,\label{eq:funcQ}
\end{equation}
this leads immediately to the Bethe equations (\ref{eq:BAE}).

For spin $1/2$ chains it has been observed \cite{YaNZ06,FrSW08}, that the
functional equations such as (\ref{eq:GrTQGleichung}) hold both in the case of
diagonal or quasi-diagonal \emph{and} in the generic off-diagonal boundary
conditions: there, only the eigenvalues of the boundary matrices enter the
equation explicitly while the deviation from constraints such as
(\ref{eq:LRconstraint}) in the non-diagonal case changes the asymptotic
behaviour of its solution.  This leads to non-polynomial solutions $q(u)$ to
the corresponding difference equations and therefore Bethe like equations are
not easily obtained. 

Based on this observation we propose that the eigenvalues of the super
transfermatrix (\ref{eq:Supertrans}) satisfy Eq. (\ref{eq:GrTQGleichung}) with
$f(u)$ parametrized by the eigenvalues of the generic boundary matrices
$K_\pm(u)$ as in (\ref{eq:funcF}).
We have verified this hypothesis for small system sizes where we are able to
explicitly construct the super transfermatrix as a square even super matrix of
corresponding finite dimension.  Taking into account the peculiarities arising
from grading as well as the nilpotency of Grassmann generators, it is
perfectly possible to perform an exact diagonalization by the use of computer
algebra systems.  For chains with up to $N=6$ sites we have computed the
eigenvalues for the most general boundary matrices $K_{-}(u)$ and
$K_{+}(u)$ and found that the functional equation (\ref{eq:GrTQGleichung})
is indeed satisfied.  Unlike the situation for spin $1/2$ chains, however, the
functions $q(u)$ are still polynomial as in (\ref{eq:funcQ}) which allows to
compute the eigenvalues by solving the Bethe equations (\ref{eq:BAE}) for
generic boundary conditions!

As an simple example we consider a system with just one site, i.e. $N=1$: the
exact diagonalization of the corresponding super transfermatrix yields the two
eigenvalues
\begin{equation}
 \Lambda^\pm(u)=-\frac{2u}{u^2-1}\klammer{1+a_{+}+u(u\pm
   1)\eckklammer{a_{+}+a_{-}(1+a_{+})}}\, . 
\end{equation}
On the other hand, assuming that the eigenvalues satisfy
(\ref{eq:GrTQGleichung}) with polynomial $q(u)$ (\ref{eq:funcQ}) we can
determine the values of the parameters $v_{\ell}$ from the requirement, that
$\Lambda(u)$ has vanishing residues at the poles at $u=v_{\ell}$ and
$u=-1-v_{\ell}$.  For $M=0$ we immediately obtain $\Lambda^+(u)$ while for
$M=1$ we find
\begin{equation}
  v_{1} =
  -\frac{1}{2}\gklammer{1\pm \sqrt{\frac{a_{-}+
        a_{+}(a_{-}-3)-4}{a_{+}+a_{-}(1+a_{+})}}}  
\end{equation}
and thereby recover the second eigenvalue $\Lambda^-(u)$.
%
%%%%%%%%%%%%%%%%%%%%%%%%%%%%%%%%%%%%%%%%%%%%%%%%%%%%%%%%%%%%%%%%%%%%%%
\section{Summary and Conclusion}
In this paper we have studied $\mathfrak{gl}(1|1)$-symmetric super chains of
free fermions subject to generic non-diagonal -- in general Grassmann valued
-- boundary fields breaking the $U(1)$ particle number conservation of the
bulk system.  The boundary conditions could be embedded into the reflection
algebra formalism resulting in quantum integrable models.  For the solution of
the spectral problem we have applied the graded algebraic Bethe ansatz for a
class of boundary conditions satisfying a constraint (\ref{eq:LRconstraint}).
In these cases both the eigenvalues and the eigenstates of the super
transfermatrix are obtained by the action of creation operators on a suitably
chosen reference state.  For generic boundary conditions such a vacuum state
could not be constructed.  Motivated by recent findings for spin chains
without grading we have proposed the hypothesis that the eigenvalues can still
be obtained from Bethe equations and verified this conjecture for small system
sizes using numerical methods.  In this case, however, it is not clear how the
eigenstates are parametrized by the Bethe roots.

Although the case of $\mathfrak{gl}(1|1)$-symmetric super chains is particular
simple since the resulting hamiltonian describes free particles, our results
indicate that it may be easier to deal with non-diagonal boundary fields in
integrable super spin chains than in models without grading.  A straight
forward extension is to the $q$-deformation of the system presented here.
Non-diagonal solutions to the reflection equations for the corresponding
small-polaron model have been constructed in the past \cite{GuFS99,WaFG00}.
Studies of the spectral problem for these chains, however, have been
restricted to the diagonal case.

%%%%%%%%%%%%%%%%%%%%%%%%%%%%%%%%%%%%%%%%%%%%%%%%%%%%%%%%%%%%%%%%%%%%%%
%
\begin{acknowledgments}
  We thank A. Seel for numerous discussions.
  This work has been supported by the Deutsche Forschungsgemeinschaft under
  grant no.\ Fr~737/6. 
\end{acknowledgments}

\appendix
\section{Superalgebras and -matrices}
\subsection{General linear Lie Superalgebras}
\label{sec:gl11}
Let $N,m,n \in \mathbb{N}$ and $\gklammer{e_j,_\alpha^{~\beta}}_{\alpha ,\beta = 1,\dots ,m+n}^{j=1,\dots ,N}$ be a homogeneous basis of an associative superalgebra, subject to the commutation relations
\begin{equation}
 \Skommut{e_j,_\alpha^{~\beta}}{e_k,_\gamma^{~\delta}}=\delta_{jk}\klammer{\delta_\gamma^\beta e_j,_\alpha^{~\delta}
  -(-1)^{p\klammer{e_j,_\alpha^{~\beta}}p\klammer{e_k,_\gamma^{~\delta}}}\delta_\alpha^\delta e_j,_\gamma^{~\beta}}\, ,
 \label{eq:superCommRel}
\end{equation}
whereas $\Skommut{X}{Y} \equiv XY-(-1)^{p(X)p(Y)}YX$ denotes the so-called \emph{super commutator} and $p(X)$ gives the parity of a homogeneous element $X$ of the superalgebra, that is
\begin{equation}
 p(X)=\begin{cases}
       0 & \text{if~} X \text{~is an element of the even subspace, or}\\
       1 & \text{if~} X \text{~is an element of the odd subspace} \, .\\
      \end{cases}
 \label{eq:parity_map}
\end{equation}
Considering the super commutator as a generalized Lie product, the generators $e_j,_\alpha^{~\beta}$ constitute the Lie superalgebra $\mathfrak{gl}(m|n)$. We restrict ourselves to the special case $m=n=1$. By identifying
\begin{equation}
 c_j^{\phantom{\dagger}} \equiv e_j,_1^{~2}\quad,\quad \hermit{c}_j \equiv e_j,_2^{~1}
 \quad,\quad n_j^{\phantom{\dagger}} \equiv \hermit{c}_j c_j^{\phantom{\dagger}} \equiv e_j,_2^{~2}
 \quad\text{and}\quad \bar{n}_j^{\phantom{\dagger}} \equiv c_j^{\phantom{\dagger}}\hermit{c}_j = 1-n_j^{\phantom{\dagger}} \equiv e_j,_1^{~1}
 \label{eq:gl11fermi}
\end{equation}
we find $\mathfrak{gl}(1|1)$ to be the algebra $\mathcal{F}$ of operators $\hermit{c}_j$ and $c_j^{\phantom{\dagger}}$ creating and annihilating spinless fermions on a one-dimensional lattice respectively, $j$ being the site index. In this case the even subspace is spanned by $n_j^{\phantom{\dagger}}$ and $\bar{n}_j^{\phantom{\dagger}}$ while $\hermit{c}_j$ and $c_j^{\phantom{\dagger}}$ span the odd subspace.

For a more detailed introduction to the construction of superalgebras on graded vector spaces, we refer to \cite{Corn89}, \cite{GoMu98} and section 12.3 in \cite{HUBBARD}.
\subsection{Grassmann algebras}
\label{sec:grassm}
Grassmann numbers, being the elements of a Grassmann algebra, are one of the
key ingredients in the formulation of non-diagonal boundary conditions for
super spin chains. The $\mathcal{N}\in\mathbb{N}$ generators of a Grassmann
algebra will be denoted by $\grass_1,\grass_2,\dots,\grass_{\mathcal{N}}$ and
in accordance with \cite{Corn89} we define a product between them such that
for all $j,k,l = 1,2,\dots,\mathcal{N}$ 
\begin{enumerate}
	\item the product is associative,
	      \begin{equation}
	        (\grass_j\grass_k)\grass_l = \grass_j(\grass_k\grass_l)\, ,
	        \label{eq:Grass1}
	      \end{equation}
	\item any two generators mutually anticommute,
	      \begin{equation}
	        \grass_j\grass_k = -\grass_k\grass_j\, ,
	        \label{eq:Grass2}
	      \end{equation}
	\item and each non-zero product
	      \begin{equation}
	        \grass_{j_1}\grass_{j_2}\dots\grass_{j_r}\quad , 1\leq r\leq\mathcal{N}
	        \label{eq:Grass3}
	      \end{equation}
	      involving $r$ generators is linearly independent of products involving less than $r$ generators. In particular, this means that Grassmann generators $\grass_j$ have \emph{no inverse}.
\end{enumerate}
For consistency reasons it is customary to supplement the set of generators by an identity $1$ with the defining properties $1\cdot 1=1$ and $1\grass_j=\grass_j 1 = \grass_j$. Using multi-index notation, each product of $|\mu |$ generators can be written as $\grass_\mu \equiv \grass_{j_1}\grass_{j_2}\dots\grass_{j_{|\mu |}}$, whereas $\mu = \gklammer{j_1,j_2,\dots,j_{|\mu |}}$ is an, without loss of generality, ascendingly ordered set of natural numbers $1\leq j_n\leq \mathcal{N}$. The identity may be incorporated by setting $\grass_\emptyset \equiv 1$. Finally, this enables us to express every Grassmann number $G$ as a linear combination of generator products $\grass_\mu$ with complex coefficients $G^\mu$,
\begin{equation}
 G = G^\mu\grass_\mu\, .
 \label{eq:GrassExpansion}
\end{equation}
Here the summation is to be carried out over all multi-indices $\mu$. In the following text this complex Grassmann algebra with $\mathcal{N}$ generators will be labeled $\mathbb{C}\boldsymbol{G}_{\mathcal{N}}$. We impose a convenient grading, setting
\begin{equation}
 p\klammer{\grass_\mu} \equiv |\mu |\,\operatorname{mod} 2\, .
\end{equation}
The complex conjugation of a Grassmann number $G$ is given by the complex conjugation of the linear coefficients in (\ref{eq:GrassExpansion}), i.e. $G^* \equiv (G^\mu)^*\grass_\mu$. Moreover we define the adjoint $\sAd{G}$ of a Grassmann number $G$ by
\begin{equation}
 \sAd{G} \equiv (-\bi)^{p(\grass_\mu)} (G^\mu)^*\grass_\mu\, .
 \label{eq:GrassAdjoint}
\end{equation}
\subsection{Super matrices}
Just like the elements of the above superalgebras, super matrices are graded objects. Here we will only make use of square \emph{even} invertable super matrices $\mbs{M}$, having the partitioning
\begin{equation}
 \mbs{M}=
 \begin{pmatrix}
  \mbs{A} & \mbs{B}\\
  \mbs{C} & \mbs{D}
 \end{pmatrix}\, ,
 \label{eq:MatrixPartition}
\end{equation}
such that all entries of the submatrices $\mbs{A}$ and $\mbs{D}$ are \emph{even} elements of a superalgebra, whereas all entries of the submatrices $\mbs{B}$ and $\mbs{C}$ are \emph{odd} elements of the same superalgebra. We define convenient analogs to the usual matrix operations. The \emph{supertrace} is given by
\begin{equation}
 \str{\mbs{M}}\equiv\spur{\mbs{A}}-\spur{\mbs{D}}\, .
 \label{eq:superTrace}
\end{equation}
In contrast to the ordinary matrix transposition, the \emph{super transposition} $(~)^\st$ is not an involution. Therefore, we have an additional \emph{inverse super transposition} $(~)^\ist$,
\begin{equation}
 \mbs{M}^{\st} \equiv \begin{pmatrix}
                        \mbs{A}^\trans & \mbs{C}^\trans \\
                       -\mbs{B}^\trans & \mbs{D}^\trans
                      \end{pmatrix}\quad ,\quad
 \mbs{M}^{\ist} \equiv \begin{pmatrix}
                         \mbs{A}^\trans & -\mbs{C}^\trans \\
                         \mbs{B}^\trans & \mbs{D}^\trans
                       \end{pmatrix}\, .
 \label{eq:superTrans}
\end{equation}
If the underlying superalgebra is $\mathbb{C}\boldsymbol{G}_{\mathcal{N}}$ there are two more important operations, namely the \emph{adjoint} operation
\begin{equation}
 \hermit{\mbs{M}} \equiv \begin{pmatrix}
                           (\sAd{\mbs{A}})^\trans & (\sAd{\mbs{C}})^\trans\\
                           (\sAd{\mbs{B}})^\trans & (\sAd{\mbs{D}})^\trans
                         \end{pmatrix}\, ,
 \label{eq:superAdjoint}
\end{equation}
where $\sAd{\mbs{A}}$ is defined by entrywise application of (\ref{eq:GrassAdjoint}), and the multiplication of a super matix by a Grassmann number $G$ of definite parity,
\begin{equation}
   G\cdot\mbs{M}\equiv\begin{pmatrix}
                       G\,\eins_{\dim A} & 0 \\
                       0        & (-1)^{p(G)}G\,\eins_{\dim D}
                      \end{pmatrix}
                      \begin{pmatrix}
                       \mbs{A} & \mbs{B} \\
                       \mbs{C} & \mbs{D}
                      \end{pmatrix}\, .
  \label{eq:MultiplGrass}
\end{equation}

%%%%%%%%%%%%%%%%
%              %
%  REFERENCES  %
%              %
%%%%%%%%%%%%%%%%

\end{document}